\documentclass[runningheads]{llncs}
\usepackage{graphicx}

\usepackage{hyperref}
\usepackage{xcolor}

\usepackage[title]{appendix}
\usepackage{url}
\usepackage{booktabs}
\usepackage{amsmath}
\usepackage{todonotes}
\usepackage{xspace}
\def\systemname#1{\textsf{#1}\xspace}
\newcommand{\lc}{\systemname{leanCoP}}
\newcommand{\rlc}{\systemname{rlCoP}}
\newcommand{\plc}{\systemname{plCoP}}
\makeatletter
\renewcommand\section{\@startsection{section}{1}{\z@}%
                       {-12\p@ \@plus -4\p@ \@minus -4\p@}%
                       {8\p@ \@plus 4\p@ \@minus 4\p@}%
                       {\normalfont\large\bfseries\boldmath
                        \rightskip=\z@ \@plus 8em\pretolerance=10000 }}
\makeatother

\begin{document}

\title{Prolog Technology\\ Reinforcement Learning Prover\thanks{
    ZZ was supported by the European Union, co-financed by the European Social Fund
    (EFOP-3.6.3-VEKOP-16-2017-00002) and the Hungarian National Excellence Grant
    2018-1.2.1-NKP-00008. JU and CB were funded by the \textit{AI4REASON} ERC Consolidator grant nr.
    649043, the Czech project AI\&Reasoning CZ.02.1.01/0.0/0.0/15\_003/0000466
    and the European Regional Development Fund.
  }\\
         {\small (System Description)}
}
\author{
  Zsolt Zombori\inst{1,2} \and
  Josef Urban\inst{3} \and
  Chad E. Brown\inst{3}
}

\authorrunning{Z. Zombori \and J. Urban \and C. Brown}
\institute{
  Alfr\'{e}d R\'{e}nyi Institute of Mathematics, Budapest, Hungary
\and 
  E{\"o}tv{\"o}s Lor\'{a}nd University, Budapest, Hungary
\and
  Czech Technical University of Prague, Czechia
}

\maketitle              
\begin{abstract}
  We present a reinforcement learning toolkit for experiments with
  guiding automated theorem proving in the connection calculus.  The core of
  the toolkit is a compact and easy to extend Prolog-based automated
  theorem prover called \plc. \plc builds on the \lc Prolog
  implementation and adds learning-guided Monte-Carlo Tree Search as
  done in the \rlc
  system. Other components include a
  Python interface to \plc and
  machine learners, and
  an external proof checker that verifies the validity of \plc
  proofs. The toolkit is evaluated on two benchmarks and we
  demonstrate its extendability by two additions: (1)
  guidance is extended to reduction steps and (2) the standard \lc
  calculus is extended with rewrite steps and their learned guidance.
  We
  argue that the Prolog setting is suitable
  for combining statistical 
and symbolic learning methods. 
The complete toolkit is publicly released.

\keywords{Automated theorem proving  \and Reinforcement Learning \and Logic Programming \and Connection Tableau calculus}
\end{abstract}

\section{Introduction}

Reinforcement learning (RL)~\cite{sutton1998reinforcement} is an area
of Machine Learning (ML) that has been responsible for some of the
largest recent AI
breakthroughs~\cite{SilverHMGSDSAPL16,silver2017mastering,abs-1712-01815,thinking_fast_and_slow}.
RL develops methods that advise agents to choose from multiple actions
in an environment with a delayed reward. This fits many settings in
Automated Theorem Proving (ATP), where many inferences are often
possible in a particular search state, but their relevance only
becomes clear when a proof is found.

Several learning-guided ATP systems have 
been developed that interleave proving
with supervised learning from proof searches \cite{enigma,ChvalovskyJ0U19,DBLP:conf/tableaux/GoertzelJU19,deep_guidance,tactictoe,rlcop,femalecop,malecop,holist}. In
the saturation-style setting used by ATP systems like E~\cite{Sch02-AICOMM} and Vampire~\cite{Vampire}, direct learning-based selection
of the most promising given clauses leads already to large
improvements~\cite{JakubuvU19}, without other changes to the
proof search procedure.

The situation is different in the connection tableau~\cite{LetzS01} setting, where
choices of actions rarely commute, and backtracking is very
common. This setting resembles games like Go, where Monte-Carlo tree
search~\cite{uct} with reinforcement learning
used for action selection (\emph{policy}) and state evaluation (\emph{value}) has
recently achieved superhuman performance. First experiments with the
\rlc system in this setting have been encouraging~\cite{rlcop},
achieving more than 40\% improvement on a test set after training on a
large corpus of Mizar problems.

The connection tableau setting is attractive also because of its
simplicity, leading to very compact Prolog implementations such as
\lc~\cite{leancop}. Such implementations are easy to modify and extend in various
ways~\cite{DBLP:conf/cade/Otten14,DBLP:conf/cade/Otten08}. This is particularly interesting for machine learning
research over reasoning corpora, where automated learning and addition
of new prover actions (tactics, inferences, symbolic decision
procedures) based on previous proof traces seems to be a large
upcoming topic. Finally, the proofs obtained in this setting are easy
to verify, which is important whenever automated self-improvement is involved.

The goal of the work described here is to develop a reinforcement
learning toolkit for experiments with guiding automated theorem
proving in the connection calculus.  The core of the toolkit
(Section~\ref{sec:plcop}) is a compact and easy to extend Prolog-based
automated theorem prover called \plc.
\plc builds on the \lc Prolog implementation and adds learning-guided
Monte-Carlo Tree Search as done in the \rlc~\cite{rlcop} system. Other
components include a Python interface to \plc and state-of-the-art
machine learners and an external proof checker that verifies the
validity of the \plc proofs. The proof checker has proven useful in
discovering bugs during development. Furthermore, it is our long term
goal to add new prover actions automatically, where proof checking
becomes essential.

Prolog is
traditionally associated with ATP research, and it has been used for a number of Prolog
provers~\cite{pttp,leantap,leancop,dlog}, as well as for rapid ATP
prototyping, with core methods like unification for free. Also, Prolog is the basis
for Inductive Logic Programming (ILP)~\cite{MuggletonR94} style systems and a natural choice for
combining such symbolic learning methods with machine learning for ATP systems, which
we are currently working on~\cite{ZU20}.
In more detail, the main contributions are:
\begin{enumerate}
  \item We provide an open-source Prolog implementation of \rlc, called \plc,
    that uses the SWI-Prolog~\cite{swi_prolog} environment.
  \item We extend the guidance of \lc to reduction steps involving unification.
  \item We extend \lc with rewrite steps while keeping the original
    equality axioms. This demonstrates the benefit of adding
    a useful but redundant inference rule, with its use controlled by the learned guidance. 
  \item We provide an external proof checker that certifies the validity of the proofs.
  \item The policy model of \rlc is trained using Monte Carlo search trees of
    all proof attempts. We show, however, that this introduces a lot of noise,
    and we get significant improvement by limiting policy training data to
    successful theorem proving attempts.
  \item Policy and value models are trained in \rlc using a subset of
    the Monte Carlo search nodes, called \emph{bigstep
      nodes}. However, when a proof is found, not all nodes leading to
    the proof are necessarily bigstep nodes. We make training more
    efficient by explicitly ensuring that all nodes leading to proofs
    are included in the training dataset.
  \item The system is evaluated in several iterations on two
    MPTP-based~\cite{Urban06} benchmarks, showing large performance increases
    thanks to learning (Section~\ref{sec:eval}). We also improve upon \rlc with $12\%$ and $7\%$ on these benchmarks.
\end{enumerate}

\section{Prolog Technology Reinforcement Learning Prover}
\label{sec:plcop}

The toolkit is available at our repository.\footnote{\url{https://github.com/zsoltzombori/plcop}}
Its core is our \plc connection prover based on the \lc 
implementation and inspired by \rlc.
\lc~\cite{leancop} is a compact theorem
prover for first-order logic, implementing connection tableau
search. The proof search starts with a \emph{start clause} as a
\emph{goal} and proceeds by building a connection tableau by applying
\emph{extension steps} and \emph{reduction steps}.
\lc uses
iterative deepening 
to ensure
completeness. This is removed in \rlc and \plc and
learning-guided Monte-Carlo Tree Search (MCTS)~\cite{mcts}
is used instead. 
Below, we explain the main ideas and parts of the system.

  \vspace{-3mm}
\paragraph{\bf{Monte Carlo Tree Search (MCTS)}}
\label{sec:mcts}
is a
search algorithm for sequential decision
processes.  MCTS builds a tree whose nodes are states of the
process, and edges represent sequential decisions.
Each state (node) yields some
reward. The aim of the search algorithm is to find trajectories (branches
in the search tree) that yield high accumulated reward.
The search starts from a single root node (\emph{starting state}), and new nodes
are added iteratively. In each node $i$, we maintain the number of
visits $n_i$, the total reward $r_i$, and its prior probability
$p_i$ given by a learned \emph{policy} function. 
Each iteration, also called \emph{playout}, starts with the addition of a new
leaf node. This is done by recursively selecting a child
that maximizes the standard UCT~\cite{uct} formula (\ref{eq:uct}), until a leaf is reached.
In~(\ref{eq:uct}), $N$ is the number of visits of the parent, and $cp$ is a parameter that
determines the balance between nodes with high value (exploitation) and rarely
visited nodes (exploration). Each leaf is given an initial value, which is typically
provided by a learned \emph{value} function. Next, ancestors are updated: visit
counts are increased by 1 and value estimates are increased by the value of
the new node.
The value and policy functions are learned in AlphaGo/Zero, \rlc and \plc.
    \vspace{-2mm}
\begin{equation}\label{eq:uct}
  \mbox{UCT}(i) = \frac{r_i}{n_i} + cp \cdot p_i \cdot \sqrt{\frac{ln N}{n_i}}
\end{equation}

  \vspace{-5mm}
\paragraph{{\bf MCTS for Connection Tableau:}}
Both \rlc and \plc use the DAgger~\cite{dagger} meta-learning algorithm to learn the policy and value functions.  
DAgger interleaves ATP runs based on the current policy and value (\emph{data collection phase}) with a \emph{training phase}, in which these functions
are updated to fit the collected data. Such iterative interleaving of proving and learning has also been used successfully
in ATP systems such as MaLARea~\cite{US+08} and ENIGMA~\cite{JakubuvU19}.
During the proof search \plc builds a Monte Carlo tree for each training
problem. Its nodes are the proof states (partial tableaux), and edges represent
inferences.
A branch leading to a node with a closed tableau is a valid
proof. Initially, \plc uses simple heuristic value and policy functions, later to be
replaced with learned guidance.
To enforce deeper exploration, \rlc and \plc perform a \emph{bigstep} after a
fixed number of playouts: the starting node of exploration is moved one level
down towards the child with the highest value (called \emph{bigstep
  node}). Later MCTS steps thus only extend the subtree under the bigstep node.

Training data for policy and value learning is extracted from the
tableau states of the bigstep nodes. The value model gets features
from the current goal and path, while the policy model also receives
features of the given action. Alternatively, one could extract policy
features from the state produced by the given action. We briefly
experimented with this, but the results were somewhat weaker. We
believe it is because the action space is much smaller than the state
space, and it is often easier to select the best action than to compare
states.

We use term walks of length up to 3 as
the main features. Both \rlc and \plc add also several more specific
features.\footnote{Number of open goals, number of symbols in them,
  their maximum size and depth, length of the current path, and two
  most frequent symbols in open goals.} The resulting sparse feature vectors
are compressed to a fixed size $d$: vector
$f$ is compressed to $f'$, so that $f'_i = \sum_{\{j | j \bmod d =
  i\}} f_j$.  For learning value, each bigstep node is assigned a
label of 1\footnote{A discount factor of 0.99 is applied to positive
  rewards to favor shorter proofs.} if it leads to a proof and 0
otherwise. The policy model gets a target probability for each edge
based on the relative frequency of the corresponding child. Both \rlc
and \plc use fast gradient boosted trees (XGBoost~\cite{xgboost}) for
guidance. Running the training concludes one iteration of
the DAgger method. More details about our policy and value functions
are provided in Appendix~\ref{app:policy_value}.

  \vspace{-2mm}
\paragraph{{\bf Prolog Implementation of \plc:}}
To implement MCTS, we modify \lc so that the Prolog stack is explicitly
maintained and saved in the Prolog database using assertions after each
inference step. This is done in the \texttt{leancop\_step.pl}
submodule, described in Appendix~\ref{app:leancop_step}.  This setup makes it
possible to interrupt proof search and later continue at any previously
visited state, required to interleave prover steps with Monte Carlo tree
steps, also implemented in Prolog. The main MCTS code is explained in
Appendix~\ref{app:mcts}. The MCTS search tree is stored in destructive Prolog hashtables.\footnote{The \texttt{hashtbl} library of SWI by G. Barany -- \url{https://github.com/gergo-/hashtbl}.}
These are necessary for efficient updates of the nodes statistics.  The training data after each proof run is exported from the
MCTS trees and saved for the XGBoost learning done in Python.

To guide the search, the trained XGBoost policy and value functions are accessed efficiently via the C foreign
language interface of SWI-Prolog. This is done in 70 lines of C++ code, using the
SWI C++ templates and the XGBoost C++ API. The main foreign predicate
\texttt{xgb:predict} takes an XGBoost predictor and a feature list and returns the
predicted value. A trained model performs 1000000 predictions in 19 sec in SWI. To
quantify the total slowdown due to the guidance, we ran \plc with 200000 inference
step limit and a large time limit (1000s) 
on the
M2k dataset (see Section~\ref{sec:eval}) with and without guidance.
The average execution time ratio on problems unsolved in both cases is
$2.88$, i.e., the XGBoost guidance roughly triples the execution time.
Efficient feature collection in \plc is implemented
using declarative association
lists (the \texttt{assoc} SWI library) implemented as AVL trees. 
Insertion, change, and retrieval is $O(log(N))$.  We also use destructive
hashtables for caching the features already computed.  Compared to \rlc, we
can rely on SWI's fast internal hash function \texttt{term\_hash/4} that only
considers terms to a specified depth. This all contributes to \plc's smaller size.

  \vspace{-2mm}
\paragraph{\bf Guiding Reduction Steps:}
The connection tableau calculus has two steps: 1)
extension that replaces the current goal with a new set of goals and
2) reduction that closes off the goal by unifying it with a literal on
the path. \rlc applies reduction steps eagerly,
which can be harmful by triggering
some unwanted unification.
Instead, \plc 
lets the guidance system learn when to apply reduction.  Suppose the
current goal is $G$. 
An input clause $\{H, B\}$, s.t. $H$ is a literal that unifies with
$\neg G$ and $B$ is a clause, yields an extension step, represented as
$ext(H, B)$, while a literal $P$ on the path that unifies with $\neg
G$ yields a reduction step, represented as $red(P)$.  The symbols
$red$ and $ext$ are then part of the standard feature representation.

    \vspace{-3mm}
\paragraph{\bf Limited Policy Training:}
\rlc extracts
policy training data from the child visit frequencies of the bigstep
nodes. We argue, however, that node visit frequencies may not be
useful when no proof was found, i.e., when no real reward was
observed. A frequently selected action that did not lead to proof
should not be reinforced. Hence, \plc only extracts policy training data
when a proof was found. Note that the same is not true for value
data. If MCTS was not successful, then bigstep nodes are given a value
of 0, which encourages exploring elsewhere.

\vspace{-3mm}
\paragraph{\bf Training from all proofsteps:}
Policy and value models are trained in \rlc using bigstep
nodes. However, when a proof is found, not all nodes leading to the
proof are necessarily bigstep nodes. We make training more efficient
by explicitly ensuring that all nodes leading to proofs are included
in the training dataset.

    \vspace{-3mm}
\paragraph{\bf (Conditional) Rewrite steps:}
\plc extends \lc with rewrite steps that can handle equality
predicates more efficiently. Let $t|_p$ denote the subterm of $t$ at
position $p$ and $t[u]_p$ denote the term obtained after replacing in
$t$ at position $p$ by term $u$. Given a goal $G$ and 
an input clause $\{X = Y, B\}$, 
s.t. for some position $p$ there is a
substitution $\sigma$ such that $G|_p\sigma = X\sigma$, the rewrite
step changes $G$ to $\{G[Y]_p\sigma, \lnot B \sigma\}$.  Rewriting is
allowed in both directions, i.e., the roles of $X$ and $Y$ can be
switched.~\footnote{The rewrite step could probably be made more
powerful by ordering equalities via a term ordering. However, we
wanted to use as little human heuristics as possible and let the
guidance figure out how to use the rewrite steps. As our long-term
goal is to learn new inference steps from proof traces, in those
situations it will be important to automatically figure out their best
usage.} This is a valid and well-known inference step, which can make
proofs much shorter. On the other hand, rewriting can be simulated by
a sequence of extension steps.  We add rewriting without removing the
original congruence axioms, making the calculus redundant. We find,
however, that the increased branching in the search space is
compensated by learning since we only explore branches that are
deemed ``reasonable'' by the guidance.

    \vspace{-3mm}
\paragraph{\bf Proof Checking:}
After \plc generates a proof, the \texttt{leancheck} program included in the toolkit
does an independent verification. Proofs are first translated into the
standard \lc proof format, with the exception of rewriting steps (see below).
The converted output from \plc includes the problem's input clauses
and a list of clauses that contributed to the proof. The proof
clauses are either input clauses, their instances, or extension
step clauses.  Proof clauses may have remaining uninstantiated (existential)
variables.  However, for proof checking, we can consider these to be new constants,
and so we consider each proof clause to be ground.  To
confirm we have a proof, it suffices to verify two assertions:
\vspace{-3mm}
\begin{enumerate}
\item The proof clause alleged to be an input clause, or its instance is subsumed by the corresponding input clause (as identified in the proof by a label).
\item The set of such proof clauses forms a propositionally unsatisfiable set.
\end{enumerate}
\vspace{-3mm}
These two properties are not hard to check, but each involves a slight complication. 
Each proof clause alleged to be an instance of an input clause
is reported as a clause $B$, a substitution $\theta$ and 
a reference to an input clause $C$.
Optimally, the checker should verify that each
literal in $\theta(C)$ is in $B$, so that $\theta(C)$ propositionally
subsumes $B$. In many cases this is what the checker verifies.
However, Prolog may rename variables so that the domain of $\theta$
no longer corresponds to the free variables of $C$.
In this case, the checker computes a renaming $\rho$ such
that $\theta(\rho(C))$ propositionally subsumes $B$.
(Note that this $\rho$ may not be unique.
Consider $C = \{p(X),p(Y)\}$, $B = \{p(c),p(c)\}$
and $\theta = W \mapsto c, Z\mapsto c$.)
We could alternatively use first-order matching to check
if $C$ subsumes $B$. This would guarantee a correct proof exists,
although it would accept proofs for which the reported $\theta$
gives incorrect information about the intended instantiation.
For the second property, we verify the propositional unsatisfiability of this ground clause set using PicoSat~\cite{Biere2008}.
There is a conceptual issue in that {\plc} is
proving a theorem given in disjunctive normal form
while PicoSat is refuting a set of clauses.
For example, the {\plc} clause for transitivity of equality appears
as $X=Y,Y=Z,X\not=Z$ instead of $X\not=Y,Y\not=Z,X=Z$.
When translating this to PicoSat we swap the polarities of
literals and ask about unsatisfiability of the resulting
conjunction of disjuncts. For example, each clause given
to PicoSat translated from an instance of transitivity would
have two negative literals and one positive literal.
Technically swapping the polarities
is not necessary since unsatisfiability is invariant under
such a swap. However, there is no extra cost since
the choice of polarity is made when translating from
the {\plc} proof to an input for PicoSat.
An example is given in Appendix~\ref{app:check}.

In case the proof uses rewriting steps, further preprocessing is required.
{\plc} reports rewriting steps by referencing the relevant input clause (with an equational literal)
along with its substitution and by giving the goal before rewriting,
the goal after rewriting, the equation used, and the side literals of the instantiated input clause.
Using this information, {\texttt{leancheck}} replaces the rewriting step with
finitely many instances of equational axioms (reflexivity, symmetry, transitivity,
and congruence) and proceeds as if there were no rewriting steps.

\section{Evaluation}
\label{sec:eval}
We use two datasets for evaluation. The first is the \emph{M2k}
benchmark that was introduced in \cite{rlcop}. The M2K dataset is a
selection of 2003 problems~\cite{m2k} from the larger Mizar40
dataset~\cite{mizar40}, which consists of 32524 problems from the
Mizar Mathematical Library that have been proven by several
state-of-the-art ATPs used with many strategies and high time limits
in the Mizar40 experiments~\cite{KaliszykU13b}. Based on the proofs,
the axioms were ATP-minimized, i.e., only those axioms were kept that
were needed in any of the ATP proofs found. This dataset is by
construction biased towards saturation-style ATP systems. To have an
unbiased comparison with state-of-the-art saturation-style ATP systems
such as E, we also evaluate the systems on the bushy (small) problems
from the MPTP2078 benchmark~\cite{abs-1108-3446}, which contains just
an article-based selection of Mizar problems, regardless of their
solvability by a particular ATP system.

We report the number of proofs found using a 200000 step inference
limit.
Hyperparameters (described in Appendix~\ref{app:hyperparameters}) were
selected to be consistent with those of \rlc. Partly because a lot of
effort has already been invested in tuning \rlc, and partly because we
wanted to make sure that the effects of our most important additions
are not obfuscated by hyperparameter changes. One difference from \rlc
is that they hash features to a fixed 262139 dimensional vector while
we use a 10000 dimensional feature vector for faster computation. Over
5 iterations on the M2k dataset, this even yields a small improvement
(928 vs. 940), likely due to less overfitting. As Tables~\ref{tab:m2k}
and \ref{tab:mptp2078} show, our baseline is weaker. This is partly
because Prolog is overall somewhat slower than OCaml and partly
because not all subtleties and tuning of \rlc have been
reproduced. Nevertheless, since our main focus is to make learning
more efficient, improvement with respect to the baseline can be used
to evaluate the new features.

  \vspace{-2.5mm}
\paragraph{\bf M2k experiments:}

We first evaluate the features introduced in Section~\ref{sec:plcop}
on the M2k dataset. Table~\ref{tab:m2k} shows that both limited policy
training and training from all proofsteps yield significant
performance increase: together, they improve upon the baseline with
$31\%$ and upon \rlc with $3\%$. However, guided reduction does not
help. We found that the proofs in this dataset tend to only use
reduction on ground goals, i.e., that does not involve unification,
which indeed can be applied eagerly. The rewrite step yields a $9\%$
increase. Improved training and rewriting together improve upon the
baseline by $42\%$ and upon \rlc by $12\%$.  Overall, thanks to
the changes in training data collection, \plc shows greater
improvement during training and finds more proofs than \rlc, even
without rewriting.
Note that \rlc has been developed and tuned on M2k.
Adding ten more iterations to the best performing ({\bf combined}) version of \plc results in 1450 problems solved, which is
17.4\% better than \rlc in 20 iterations (1235).

\begin{table}[htb]
  \caption{Performance on the M2k dataset: original {\bf \rlc}, \plc ({\bf baseline}), \plc with
    {\bf guided reduction}, \plc with {\bf limited policy} training, \plc trained using {\bf all proofsteps}, \plc using the previous two {\bf improved training} \plc with 
    {\bf rewriting} and \plc with rewriting and improved training {\bf
      combined}. {\bf incr} shows the performance increase in percentages from
    iteration 0 (unguided) to the best result.}
  \label{tab:m2k}
  \centering
  \begin{tabular}{ l l l l l l l l l l l l l }
    Iteration &  0 & 1 & 2 & 3 & 4 & 5 & 6 & 7 & 8 & 9 & 10 & incr\\
    \bottomrule
    \rlc & 770 & 1037 & 1110 & 1166 & 1179 & 1182 & 1198 & 1196 & 1193 & {\bf 1212} & 1210 & $57\%$  \\

    baseline & 632 & 852 & 860 & 915 & 918 & 944 & 949 & {\bf 959} & 955 & 943 & 954 & $52\%$  \\

    guided reduction & 616 & 840 & 884 & 905 & 915 & 900 & 914 & 924 & {\bf 942} & 915 & 912 & $53\%$ \\

    limited policy & 632 & 988 & 1037 & 1071 & 1080 & 1094 & 1092 & 1101 & 1103 & {\bf 1118} & 1111 & $77\%$ \\

    all proofsteps & 632 & 848 & 930 & 988 & 986 & 1018 & 1033 & 1039 & {\bf 1053} & 1043 & 1050 & $67\%$ \\

    improved training & 632 & 975 & 1100 & 1154 & 1180 & 1189 & 1209 & 1231 & 1238 & 1243 & {\bf 1254} & {\bf 98\% } \\

    rewriting & 695 & 913 & 989 & 995 & 1003 & 1019 & 1030 & 1030 & 1033 & 1038 & {\bf 1045} & $50\%$ \\

    combined & 695 & 1070 & 1209 & 1253 & 1295 & 1309 & 1322 & 1335 & 1339 & 1346 & {\bf 1359} & $96\%$ \\
    \bottomrule
  \end{tabular}
\end{table}

\vspace{-2mm}
\paragraph{\bf MPTP2078}

Using a limit of 200000 inferences, unmodified \lc solves 612 of the
MPTP2078 bushy problems, while its OCaml version (mlcop), used as a
basis for \rlc solves 502. E solves 998, 505, 326, 319 in
\emph{auto}, \emph{noauto}, \emph{restrict}, \emph{noorder}
modes\footnote{For a description of these E configurations, see
  Table~9 of \cite{rlcop}.}  
\plc and \rlc results are summarized in
Table~\ref{tab:mptp2078}. 
Improved training and rewriting together yield $63\%$ improvement upon the baseline and $7\%$ improvement upon \rlc.
Here, it is \plc that starts better, while \rlc shows stronger
learning. Eventually, \plc outperforms \rlc, even without
rewriting. Additional ten iterations of the {\bf combined} version
increase the performance to 854 problems. This is 12\% more than
\rlc in 20 iterations (763) but still weaker than the strongest E
configuration (998). However it performs better than E with the
more limited heuristics, as well as \lc with its heuristics.

\begin{table}[htb]
    \caption{Performance on the MPTP2078 bushy dataset: original {\bf \rlc}, {\bf baseline} \plc, \plc using {\bf improved training} and {\bf combined} \plc.
          {\bf incr} shows the performance increase in percentages from iteration 0 (unguided) to the best result.}

  \label{tab:mptp2078}
  \centering
  \begin{tabular}{ l  l l l l l l l l l l l l l}
Iteration & 0 &   1 &   2 &   3 &   4 &   5 &   6 &   7 &   8 &   9 & 10 & incr \\
    \toprule
    \rlc    & 198 & 300 & 489 & 605 & 668 & 701 & 720 & 737 & 736 & 732 & {\bf 733} & {\bf 270\%} \\    

    baseline & 287 & 363 & 413 & 420 & 429 & 441 & 454 & 464 & 465 & {\bf 479} & 469 & $67\%$  \\

    improved training & 287 & 449 & 544 & 611 & 640 & 674 & 692 & 704 & 720 & 731 & {\bf 744} & $140\%$\\

    combined & 326 & 460 & 563 & 642 & 671 & 694 & 721 & 740 & 761 & 775 & {\bf 782} & $140\%$ \\

\bottomrule
\end{tabular}
\vspace{-5mm}
\end{table}

\section{Conclusion and Future Work}

We have developed a reinforcement learning toolkit for experiments
with guiding automated theorem proving in the connection calculus.
Its core is the Prolog-based \plc obtained by extending \lc with a
number of features motivated by \rlc.  New features on top of \rlc
include guidance of reduction steps, the addition of the rewrite
inference rule and its guidance, external proof checker, and
improvements to the training data selection. Altogether, \plc improves
upon \rlc on the M2K and the MPTP2078
datasets by $12\%$ and $7\%$ in ten iterations, and by $17.4\%$ and $12\%$ in twenty iterations.
The system is publicly available to the ML and
AI/TP communities for experiments and extensions.

One lesson learned is that due to the sparse rewards in theorem
proving, care is needed when extracting training data from the
prover's search traces. 
Another lesson is that new sound inference rules can
be safely added to the underlying calculus. Thanks to the guidance,
the system still learns to focus on promising actions, while the new
actions may yield shorter search and proofs for some problems. An
important part of such an extendability scheme is the independent proof checker provided.

Future work includes, e.g., the addition of neural learners, such as tree and graph neural networks~\cite{ChvalovskyJ0U19,DBLP:journals/corr/abs-1911-12073}.
An important motivation for choosing Prolog is our plan to employ
Inductive Logic Programming to learn new prover actions (as Prolog programs) from the \plc proof traces.
As the manual addition of the rewrite step already shows, such
new actions can be inserted into the proof search, and guidance can again be trained to use them efficiently.
Future work, therefore, involves AI/TP research in combining statistical and symbolic learning in this framework,
with the goal of automatically learning more and more complex actions 
similar to tactics in interactive theorem
provers. We believe this may become a very interesting AI/TP research topic facilitated by the toolkit.

\bibliographystyle{splncs04}
\bibliography{plcop,ate11}

\begin{thebibliography}{10}
\providecommand{\url}[1]{\texttt{#1}}
\providecommand{\urlprefix}{URL }
\providecommand{\doi}[1]{https://doi.org/#1}

\bibitem{abs-1108-3446}
Alama, J., Heskes, T., K\"{u}hlwein, D., Tsivtsivadze, E., Urban, J.: Premise
  selection for mathematics by corpus analysis and kernel methods. J. Autom.
  Reasoning  \textbf{52}(2),  191--213 (2014). \doi{10.1007/s10817-013-9286-5}

\bibitem{Andrews1989}
Andrews, P.B.: {On Connections and Higher-Order Logic}. Journal of Automated
  Reasoning  \textbf{5}(3),  257--291 (1989)

\bibitem{thinking_fast_and_slow}
Anthony, T., Tian, Z., Barber, D.: Thinking fast and slow with deep learning
  and tree search. CoRR  \textbf{abs/1705.08439} (2017),
  \url{http://arxiv.org/abs/1705.08439}

\bibitem{holist}
Bansal, K., Loos, S.M., Rabe, M.N., Szegedy, C., Wilcox, S.: {HOList}: An
  environment for machine learning of higher-order theorem proving (extended
  version). CoRR  \textbf{abs/1904.03241} (2019),
  \url{http://arxiv.org/abs/1904.03241}

\bibitem{leantap}
Beckert, B., Posegga, J.: leantap: Lean tableau-based deduction. Journal of
  Automated Reasoning  \textbf{15},  339--358 (1995)

\bibitem{Bibel1987}
Bibel, W.: Automated theorem proving. Artificial Intelligence, Vieweg, 2ed edn.
  (1987)

\bibitem{Biere2008}
Biere, A.: Picosat essentials. Journal on Satisfiability, Boolean Modeling and
  Computation (JSAT  \textbf{4},  75--97 (2008)

\bibitem{mcts}
Browne, C., Powley, E.J., Whitehouse, D., Lucas, S.M., Cowling, P.I.,
  Rohlfshagen, P., Tavener, S., Liebana, D.P., Samothrakis, S., Colton, S.: A
  survey of monte carlo tree search methods. IEEE Transactions on Computational
  Intelligence and AI in Games  \textbf{4},  1--43 (2012)

\bibitem{xgboost}
Chen, T., Guestrin, C.: {XGBoost}: A scalable tree boosting system. In:
  Proceedings of the 22Nd ACM SIGKDD International Conference on Knowledge
  Discovery and Data Mining. pp. 785--794. KDD '16 (2016),
  \url{http://doi.acm.org/10.1145/2939672.2939785}

\bibitem{ChvalovskyJ0U19}
Chvalovsk{\'{y}}, K., Jakubuv, J., Suda, M., Urban, J.: {ENIGMA-NG:} efficient
  neural and gradient-boosted inference guidance for {E}. In: Fontaine, P.
  (ed.) Automated Deduction - {CADE} 27 - 27th International Conference on
  Automated Deduction, Natal, Brazil, August 27-30, 2019, Proceedings. Lecture
  Notes in Computer Science, vol. 11716, pp. 197--215. Springer (2019).
  \doi{10.1007/978-3-030-29436-6},
  \url{https://doi.org/10.1007/978-3-030-29436-6\_12}

\bibitem{tactictoe}
Gauthier, T., Kaliszyk, C., Urban, J., Kumar, R., Norrish, M.: Learning to
  prove with tactics. CoRR  \textbf{abs/1804.00596} (2018),
  \url{http://arxiv.org/abs/1804.00596}

\bibitem{DBLP:conf/tableaux/GoertzelJU19}
Goertzel, Z., Jakubuv, J., Urban, J.: {ENIGMAWatch}: {ProofWatch} meets
  {ENIGMA}. In: Cerrito, S., Popescu, A. (eds.) Automated Reasoning with
  Analytic Tableaux and Related Methods - 28th International Conference,
  {TABLEAUX} 2019, London, UK, September 3-5, 2019, Proceedings. Lecture Notes
  in Computer Science, vol. 11714, pp. 374--388. Springer (2019).
  \doi{10.1007/978-3-030-29026-9},
  \url{https://doi.org/10.1007/978-3-030-29026-9\_21}

\bibitem{enigma}
Jakubuv, J., Urban, J.: {ENIGMA:} efficient learning-based inference guiding
  machine. In: Geuvers, H., England, M., Hasan, O., Rabe, F., Teschke, O.
  (eds.) Intelligent Computer Mathematics - 10th International Conference,
  {CICM} 2017, Edinburgh, UK, July 17-21, 2017, Proceedings. Lecture Notes in
  Computer Science, vol. 10383, pp. 292--302. Springer (2017).
  \doi{10.1007/978-3-319-62075-6\_20},
  \url{https://doi.org/10.1007/978-3-319-62075-6\_20}

\bibitem{JakubuvU19}
Jakubuv, J., Urban, J.: Hammering {Mizar} by learning clause guidance. In:
  Harrison, J., O'Leary, J., Tolmach, A. (eds.) 10th International Conference
  on Interactive Theorem Proving, {ITP} 2019, September 9-12, 2019, Portland,
  OR, {USA}. LIPIcs, vol.~141, pp. 34:1--34:8. Schloss Dagstuhl -
  Leibniz-Zentrum f{\"{u}}r Informatik (2019),
  \url{https://doi.org/10.4230/LIPIcs.ITP.2019.34}

\bibitem{m2k}
Kaliszyk, C., Urban, J.: {M2K} dataset,
  \url{https://github.com/JUrban/deepmath/blob/master/M2k_list}

\bibitem{mizar40}
Kaliszyk, C., Urban, J.: {Mizar40} dataset,
  \url{https://github.com/JUrban/deepmath}

\bibitem{femalecop}
Kaliszyk, C., Urban, J.: {FEMaLeCoP}: Fairly efficient machine learning
  connection prover. In: Davis, M., Fehnker, A., McIver, A., Voronkov, A.
  (eds.) Logic for Programming, Artificial Intelligence, and Reasoning - 20th
  International Conference, 2015, Proceedings. Lecture Notes in Computer
  Science, vol.~9450, pp. 88--96. Springer (2015).
  \doi{10.1007/978-3-662-48899-7},
  \url{https://doi.org/10.1007/978-3-662-48899-7\_7}

\bibitem{KaliszykU13b}
Kaliszyk, C., Urban, J.: {MizAR 40 for Mizar 40}. J. Autom. Reasoning
  \textbf{55}(3),  245--256 (2015),
  \url{{http://dx.doi.org/10.1007/s10817-015-9330-8}}

\bibitem{rlcop}
Kaliszyk, C., Urban, J., Michalewski, H., Ols{\'{a}}k, M.: Reinforcement
  learning of theorem proving. In: NeurIPS. pp. 8836--8847 (2018)

\bibitem{uct}
Kocsis, L., Szepesv{\'a}ri, C.: Bandit based monte-carlo planning. In:
  F{\"u}rnkranz, J., Scheffer, T., Spiliopoulou, M. (eds.) Machine Learning:
  ECML 2006. pp. 282--293. Springer Berlin Heidelberg, Berlin, Heidelberg
  (2006)

\bibitem{Vampire}
Kov{\'a}cs, L., Voronkov, A.: First-order theorem proving and {V}ampire. In:
  Sharygina, N., Veith, H. (eds.) CAV. LNCS, vol.~8044, pp. 1--35. Springer
  (2013)

\bibitem{LetzS01}
Letz, R., Stenz, G.: Model elimination and connection tableau procedures. In:
  Robinson, J.A., Voronkov, A. (eds.) Handbook of Automated Reasoning (in 2
  volumes), pp. 2015--2114. Elsevier and {MIT} Press (2001)

\bibitem{deep_guidance}
Loos, S.M., Irving, G., Szegedy, C., Kaliszyk, C.: Deep network guided proof
  search. In: 21st International Conference on Logic for Programming,
  Artificial Intelligence, and Reasoning (LPAR) (2017)

\bibitem{dlog}
Luk{\'{a}}csy, G., Szeredi, P.: Efficient description logic reasoning in
  prolog: The dlog system. {TPLP}  \textbf{9}(3),  343--414 (2009),
  \url{https://doi.org/10.1017/S1471068409003792}

\bibitem{MuggletonR94}
Muggleton, S., Raedt, L.D.: Inductive logic programming: Theory and methods. J.
  Log. Program.  \textbf{19/20},  629--679 (1994).
  \doi{10.1016/0743-1066(94)90035-3},
  \url{https://doi.org/10.1016/0743-1066(94)90035-3}

\bibitem{DBLP:journals/corr/abs-1911-12073}
Ols{\'{a}}k, M., Kaliszyk, C., Urban, J.: Property invariant embedding for
  automated reasoning. CoRR  \textbf{abs/1911.12073} (2019),
  \url{http://arxiv.org/abs/1911.12073}

\bibitem{DBLP:conf/cade/Otten08}
Otten, J.: {leanCoP} 2.0 and {ileanCoP} 1.2: High performance lean theorem
  proving in classical and intuitionistic logic (system descriptions). In:
  Armando, A., Baumgartner, P., Dowek, G. (eds.) Automated Reasoning, 4th
  International Joint Conference, {IJCAR} 2008, Sydney, Australia, August
  12-15, 2008, Proceedings. Lecture Notes in Computer Science, vol.~5195, pp.
  283--291. Springer (2008),
  \url{https://doi.org/10.1007/978-3-540-71070-7\_23}

\bibitem{DBLP:conf/cade/Otten14}
Otten, J.: {MleanCoP}: {A} connection prover for first-order modal logic. In:
  Demri, S., Kapur, D., Weidenbach, C. (eds.) Automated Reasoning - 7th
  International Joint Conference, {IJCAR} 2014, Held as Part of the Vienna
  Summer of Logic, {VSL} 2014, Vienna, Austria, July 19-22, 2014. Proceedings.
  Lecture Notes in Computer Science, vol.~8562, pp. 269--276. Springer (2014).
  \doi{10.1007/978-3-319-08587-6},
  \url{https://doi.org/10.1007/978-3-319-08587-6\_20}

\bibitem{leancop}
Otten, J., Bibel, W.: {leanCoP}: lean connection-based theorem proving. J.
  Symb. Comput.  \textbf{36},  139--161 (2003)

\bibitem{dagger}
Ross, S., Gordon, G., Bagnell, D.: A reduction of imitation learning and
  structured prediction to no-regret online learning. In: Gordon, G., Dunson,
  D., Dudík, M. (eds.) Proceedings of the Fourteenth International Conference
  on Artificial Intelligence and Statistics. Proceedings of Machine Learning
  Research, vol.~15, pp. 627--635. PMLR, Fort Lauderdale, FL, USA (11--13 Apr
  2011), \url{http://proceedings.mlr.press/v15/ross11a.html}

\bibitem{Sch02-AICOMM}
Schulz, S.: {E - A Brainiac Theorem Prover}. AI Commun.  \textbf{15}(2-3),
  111--126 (2002)

\bibitem{SilverHMGSDSAPL16}
Silver, D., Huang, A., Maddison, C.J., Guez, A., Sifre, L., van~den Driessche,
  G., Schrittwieser, J., Antonoglou, I., Panneershelvam, V., Lanctot, M.,
  Dieleman, S., Grewe, D., Nham, J., Kalchbrenner, N., Sutskever, I.,
  Lillicrap, T.P., Leach, M., Kavukcuoglu, K., Graepel, T., Hassabis, D.:
  Mastering the game of go with deep neural networks and tree search. Nature
  \textbf{529}(7587),  484--489 (2016),
  \url{https://doi.org/10.1038/nature16961}

\bibitem{abs-1712-01815}
Silver, D., Hubert, T., Schrittwieser, J., Antonoglou, I., Lai, M., Guez, A.,
  Lanctot, M., Sifre, L., Kumaran, D., Graepel, T., Lillicrap, T.P., Simonyan,
  K., Hassabis, D.: Mastering chess and shogi by self-play with a general
  reinforcement learning algorithm. CoRR  \textbf{abs/1712.01815} (2017),
  \url{http://arxiv.org/abs/1712.01815}

\bibitem{silver2017mastering}
Silver, D., Schrittwieser, J., Simonyan, K., Antonoglou, I., Huang, A., Guez,
  A., Hubert, T., Baker, L., Lai, M., Bolton, A., et~al.: Mastering the game of
  go without human knowledge. Nature  \textbf{550}(7676), ~354 (2017)

\bibitem{pttp}
Stickel, M.E.: A prolog technology theorem prover: Implementation by an
  extended prolog computer  \textbf{4}(4) (1988),
  \url{https://doi.org/10.1007/BF00297245}

\bibitem{sutton1998reinforcement}
Sutton, R.S., Barto, A.G.: Reinforcement learning: An introduction, vol.~1.
  Cambridge Univ Press (1998)

\bibitem{Urban06}
Urban, J.: {MPTP} 0.2: Design, implementation, and initial experiments. J.
  Autom. Reasoning  \textbf{37}(1-2),  21--43 (2006)

\bibitem{US+08}
Urban, J., Sutcliffe, G., Pudl{\'a}k, P., Vysko\v{c}il, J.: {MaLARea SG1 -
  Machine Learner for Automated Reasoning with Semantic Guidance}. In: IJCAR.
  pp. 441--456 (2008)

\bibitem{malecop}
Urban, J., Vyskocil, J., Step{\'{a}}nek, P.: {MaLeCoP}: Machine learning
  connection prover. In: Br{\"u}nnler, K., Metcalfe, G. (eds.) Automated
  Reasoning with Analytic Tableaux and Related Methods - 20th International
  Conference, {TABLEAUX} 2011, Bern, Switzerland, July 4-8, 2011. Proceedings.
  LNCS, vol.~6793, pp. 263--277. Springer (2011),
  \url{https://doi.org/10.1007/978-3-642-22119-4\_21}

\bibitem{swi_prolog}
Wielemaker, J., Schrijvers, T., Triska, M., Lager, T.: {SWI-Prolog}. Theory and
  Practice of Logic Programming  \textbf{12}(1-2),  67--96 (2012)

\bibitem{ZU20}
Zombori, Z., Urban, J.: Learning complex actions from proofs in theorem
  proving, accepted to AITP'20,
  \url{http://aitp-conference.org/2020/abstract/paper_11.pdf}

\end{thebibliography}

\appendix

\section{Proof Checking Example}
\label{app:check}
As a simple example suppose we are proving a proposition $q(a)$
under two assumptions $\forall x.p(x)$ and $\forall x.p(x)\Rightarrow q(a)$.
{\plc} will start with three input ``clauses'' $p(X)^-$, $p(Y) \lor q(a)^-$
and $q(a)$. The connection proof proceeds in the obvious way and yields
three proof clauses that are either input clauses
or instances of input clauses: $p(X)^-$, $p(X) \lor q(a)^-$ and $q(a)$.
Unification during the search makes the two variables $X$ and $Y$ the same variable $X$.
For proof checking, we now consider $X$ to be constant (in the same sense as $a$).
Switching from the point of view of proving a disjunction of conjuncts to
the point of view of refuting a conjunction of disjuncts, we see these as
three propositional clauses: $P$, $P^- \lor Q$ and $Q^-$
where $P$ stands for the atom $p(X)$ (now viewed as ground)
and $Q$ stands for $q(a)$ (also ground).
The set $\{P, P^-\lor Q, Q^-\}$ is clearly unsatisfiable.
In terms of the connection method, the unsatisfiability of this set
guarantees every path~\cite{Andrews1989,Bibel1987} 
has a pair of complementary literals.

\section{Monte Carlo Tree Search in Prolog}
\label{app:mcts}

We show the most important predicates that perform the MCTS. The code has been
simplified for readability.

We repeatedly perform playouts, which consist of three steps: 1) find the next
tree node to expand, 2) add a new child to this node and 3) update ancestor values and
visit counts:

\begin{verbatim}
mc_playout(ChildHash,ParentHash,NodeHash,FHash):-

    % get the current bigstep node (root of exploration)
    rootnode(StartId), !,

    % find node to expand
    mc_find_unexpanded(StartId,ChildHash,NodeHash,
                       ExpandId,UnexpandedActionIds),
    nb_hashtbl_get(NodeHash,ExpandId,[State,_,_,_,ChProbs]),

    State=state(_,_,_,_,_,_,Result),
    (  Result ==  1 -> Reward = 1 % we found a proof
     ; Result == -1 -> Reward = 0 % proof failed
    ;  get_largest_index(UnexpandedActionIds, ChProbs, 
                         ActionIndex),
       flag(inference_cnt, X, X+1), % increase inference count

       % we expand the child with the largest prior probability
       mc_expand_node(ExpandId,ChildHash,ParentHash,NodeHash,
                      FHash,ActionIndex,Reward)
    ),

    % update ancestor visit counts and values
    mc_backpropagate(ExpandId,Reward,ParentHash,NodeHash).
\end{verbatim}

We search for the node to expand based on the standard UCT formula:

\begin{verbatim}
% +Id: current node id
% -Id2: next node id to expand
mc_find_unexpanded(Id,ChildHash,NodeHash,
                   Id2,UnexpandedActionIds):-
    mc_child_list(Id,NodeHash,ChildHash,ChildPairs),
    nb_hashtbl_get(NodeHash,Id,[State,_,VisitCount,_,_]),
    action_count(State,ActionCount),
    length(ChildPairs,L),
    (  ActionCount == 0 -> % no valid moves
       Id2=Id, UnexpandedActionIds=[] 
    ;  mc_ucb_select_child(VisitCount,ChildPairs,NodeHash,
                           SelectedId,UCBScore),
       ( L < ActionCount,
         mc_ucb_score_unexplored(VisitCount,ActionCount,
                                 UCBUnexploredScore),
         UCBUnexploredScore > UCBScore -> %select current node
         Id2=Id, 
         missing_actions(ActionCount,ChildPairs,
                         UnexpandedActionIds)
       ;
         % we move towards the child with the highest UCB score
         mc_find_unexpanded(SelectedId,ChildHash,NodeHash,
                            Id2,UnexpandedActionIds)
       ), !
    ;  % The current node is a leaf, so we select it
       Id2=Id, 
       missing_actions(ActionCount,ChildPairs,
                       UnexpandedActionIds)
    ).
\end{verbatim}

Once the node to expand has been selected, we pick the unexplored child that
has the highest UCB score:

\begin{verbatim}
mc_expand_node(ParentId,ChildHash,ParentHash,NodeHash,FHash,
               ActionIndex,ChildValue):-
    nb_hashtbl_get(NodeHash,ParentId,[ParentState,_,_,_,ChProbs]),

    % we perform the inference step corresponding to the new child
    copy_term(ParentState,ParentState2),
    logic_step(ParentState2,ActionIndex,ChildState),

    % value estimate from the external (xgboost) model
    guidance_get_value(ChildState, FHash, ChildValue),

    % probability estimates for the children of the new node
    % from the external (xgboost) model
    guidance_action_probs(ChildState,FHash,ChChProbs),   

    % store the new node in the hash tables 
    nth0(ActionIndex, ChProbs, ChProb),
    flag(nodecount, ChildId, ChildId+1),
    nb_hashtbl_set(ChildHash,ParentId-ActionIndex,ChildId),
    nb_hashtbl_set(ParentHash,ChildId,ParentId),
    nb_hashtbl_set(NodeHash,ChildId,
                   [ChildState,ChProb,1,ChildValue,ChChProbs]).
\end{verbatim}

Finally, we update ancestor nodes after the insertion of the new leaf:

\begin{verbatim}
mc_backpropagate(Id,Reward,ParentHash,NodeHash):-
    nb_hashtbl_get(NodeHash,Id,[State,Prob,VCnt,Value,ChProbs]),
    VCnt1 is VCnt + 1, 
    Value1 is Value + Reward, 
    nb_hashtbl_set(NodeHash,Id,[State,Prob,VCnt1,Value1,ChProbs]),
    ( nb_hashtbl_get(ParentHash, Id, ParentId) -> 
      mc_backpropagate(ParentId,Reward,ParentHash,NodeHash)
    ; true
    ).
\end{verbatim}

\section{\texttt{leancop\_step.pl} module}
\label{app:leancop_step}

Below we provide the code for the most important predicates that handle \lc
inference steps in such a way that the entire prover state is explicitly
maintained. For better readability, we omit some details, mostly related to
problem loading, logging, and proof reconstruction. The \texttt{nondet\_step}
predicate takes a prover state along with the index of an extension or
reduction step and returns the subsequent state. Before returning, it
repeatedly calls the \texttt{det\_steps} predicate, which performs
optimization steps that do not involve choice (loop elimination, reduction
without unification, lemma, single action).

\begin{verbatim}
% init_pure(+File,+Settings,-NewState)
init_pure(File,Settings,NewState):-
    NewState = state(Goal,Path,Lem,Actions,Todos,Proof,Result),

    % store options
    retractall(option(_)),
    findall(_, ( member(S,Settings), assert(option(S)) ), _ ), 

    % load tptp file and store contrapositives
    {...}

    % perform any potential optimizations
    det_steps([-(#)],[],[],[],[init((-#)-(-#))],
              Goal,Path,Lem,Todos,Proof,Result),

    % collect valid moves from this state
    % valid_actions(Goal, Path, Actions).

    
:- dynamic(alt/6).
% step_pure(+ActionIndex,+State,-NewState,-SelectedAction))
step_pure(ActionIndex,State,NewState,Action0):-
    State = state(Goal0,Path0,Lem0,Actions0,Todos0,Proof0,_),
    NewState = state(Goal,Path,Lem,Actions,Todos,Proof,Result),

    nth0(ActionIndex,Actions0,Action0),

    % if there were other alternative actions, store them
    (option(backtrack), Actions0=[_,_|_] ->
         select_nounif(Action0, Actions0, RemActions0), !,
         asserta(alt(Goal0,Path0,Lem0,RemActions0,Todos0,Proof0))
     ; true
    ),

    % perform any potential optimizations
    nondet_step(Action0,Goal0,Path0,Lem0,Todos0,Proof0,
                Goal1,Path1,Lem1,Todos1,Proof1,Result1),

    % if proof search fails, pop an alternative
    ( Result1 == -1, option(backtrack), 
      pop_alternative(Goal,Path,Lem,Actions,Todos,Proof) ->
          Result=0, 
     ; [Goal,Path,Lem] = [Goal1,Path1,Lem1],
       [Todos,Proof,Result] = [Todos1,Proof1,Result1],
       valid_actions(Goal,Path,Actions)
    ).

%%% make a single proof step from a choice point
% nondet_step(Action,Goal,Path,Lem,Todos,Proof,
%             NewGoal,NewPath,NewLem,NewTodos,NewProof,Result)
% reduction step
nondet_step(red(NegL),[Lit|Cla],Path,Lem,Todos,Proof,
            NewGoal,NewPath,NewLem,NewTodos,NewProof,Result):- 
    neg_lit(Lit,NegL),
    Proof2 = {...}
    det_steps(Cla,Path,Lem,Todos,Proof2,
              NewGoal,NewPath,NewLem,NewTodos,NewProof,Result).
% extension step
nondet_step(ext(NegLit,Cla1,_),[Lit|Cla],Path,Lem,Todos,Proof,
            NewGoal,NewPath,NewLem,NewTodos,NewProof,Result):-
    neg_lit(Lit, NegLit),
    ( Cla=[_|_] ->
	  Todos2 = [[Cla,Path,[Lit|Lem]]|Todos]
     ; Todos2 = Todos
    ),
    Proof2= {...}
    det_steps(Cla1,[Lit|Path],Lem,Todos2,Proof2,
              NewGoal,NewPath,NewLem,NewTodos,NewProof,Result).


% perform steps until the next choice point (or end of proof)
det_steps([],_Path,_Lem,Todos,Proof,
          NewGoal,NewPath,NewLem,NewTodos,NewProof,Result):-
    !,
    ( Todos = [] -> % nothing to prove, nothing todo on the stack
           [NewGoal,NewPath,NewLem,NewTodos,NewProof,Result] = 
           [[success],[],[],[],Proof,1]
     ; Todos = [[Goal2,Path2,Lem2]|Todos2] -> 
           % nothing to prove, something on the stack
           det_steps(Goal2,Path2,Lem2,Todos2,Proof,NewGoal,
                     NewPath,NewLem,NewTodos,NewProof,Result)
    ).
det_steps([Lit|_Cla],Path,_Lem,_Todos,Proof,
          NewGoal,NewPath,NewLem,NewTodos,NewProof,Result):-
    member(P,Path), Lit == P, !, % loop elimination
    [NewGoal,NewPath,NewLem,NewTodos,NewProof,Result] = 
    [[failure],[],[],[],Proof,-1].    
det_steps([Lit|Cla],Path,Lem,Todos,Proof,
          NewGoal,NewPath,NewLem,NewTodos,NewProof,Result):-
    member(LitL,Lem), Lit==LitL, !, % perform lemma step
    Proof2 = [lem(Lit)|Proof],
    det_steps(Cla,Path,Lem,Todos,Proof2,
              NewGoal,NewPath,NewLem,NewTodos,NewProof,Result).
det_steps([Lit|Cla],Path,Lem,Todos,Proof,
          NewGoal,NewPath,NewLem,NewTodos,NewProof,Result):-
    neg_lit(Lit,NegLit),
    ( option(eager_reduction(1)) -> 
      member(NegL,Path), 
      unify_with_occurs_check(NegL, NegLit), ! % eager reduction
    ; member(NegL,Path), 
      NegL == NegLit, ! % reduction without unification is safe
    ),
    Ext = [NegL, NegL],
    Proof2 = [red(Ext-Ext)|Proof],
    det_steps(Cla,Path,Lem,Todos,Proof2,
              NewGoal,NewPath,NewLem,NewTodos,NewProof,Result).
det_steps(Goal,Path,Lem,Todos,Proof,
          NewGoal,NewPath,NewLem,NewTodos,NewProof,Result):-
    valid_actions(Goal,Path,Actions),
    ( option(single_action_optim),  Actions==[A] -> 
          % only a single action is available, so perform it
          nondet_step(A,Goal,Path,Lem,Todos,Proof,NewGoal,
                      NewPath,NewLem,NewTodos,NewProof,Result)
    ;Actions==[] ->             % proof failed
          [NewGoal,NewPath,NewLem,NewTodos,NewProof,Result] = 
          [[failure],[],[],[],Proof,-1]
    ; option(comp(PathLim)), \+ ground(Goal), length(Path,PLen), 
      PLen > PathLim -> % reached path limit
      [NewGoal,NewPath,NewLem,NewTodos,NewProof,Result] = 
      [[failure],[],[],[],Proof,-1]
    ;[NewGoal,NewPath,NewLem,NewTodos,NewProof,Result] = 
     [Goal,Path,Lem,Todos,Proof,0]
    ).
\end{verbatim}

\section{Policy and Value Functions}
\label{app:policy_value}
Here we describe 1) the default value and policy functions used in the
first iteration, 2) the training data extraction and 3) how the
predicted model values are used in MCTS. All these formulae are taken
directly from \rlc~\cite{rlcop} and have been highly hand-engineered.
We currently use these solutions in \plc without altering
them; however, we believe some of these decisions are worth
reconsidering.

\subsection{Value Function}

In the first iteration, the default value $V_d$ is based on the total term size of all open goals. Given $s$ with total term size of open goals $t$, its value is

\begin{equation}
V(s) = \frac{1}{1+e^{-3.7 * e^{-0.05 t} + 2.5}}
\end{equation}

After the MCTS phase, training data is extracted from the states in the
bigstep nodes. If a state $s$ is $k$ steps away from a success node, 
its target value is $0.99^k$. If none of its descendants are success
nodes, then its target value is $0$. We can then build a model using
logistic regression. However, the authors of \rlc find that the
xgboost model works better if
standard regression is used, so the target
value is first mapped into the range $[-3, 3]$. The value $V_t$ used
for model training is

$$V_t(s,k) = min(3, max(-3, log(0.99^k / (1-0.99^k))))$$

In subsequent iterations, the prediction $V_p$ of the model is mapped back to the $[0,1]$ range ($V_p'$):

$$V_p'(V_p) = \frac{1}{1+e^{-V_p}}$$

This value is further adjusted to give an extra incentive towards states with few open goals. If the state has $g$ open goals, then the final value ($V_f$) used in MCTS is

$$V_f(V_p', g) = (\sqrt{V_p'})^g$$

\subsection{Policy Function}

The default policy $P_d$ is simply the uniform distribution, i.e., if a state $s$ has $n$ valid inferences, then each action $a$ has a prior probability of 

$$P_d(n) = \frac{1}{n}$$

After the MCTS phase, training data is extracted from the (state, action)
pairs in the bigstep nodes. Target probabilities are based on relative
visit frequencies of child nodes. These frequencies are again mapped
to a range where we can do 
standard
regression. Given state $s$ with $n$
valid inferences, such that $s$ was expanded $N$ times and its $j$th child was
visited $N_j$ times, then the policy $P_t$ used for model training is

$$P_t(s,n,N,N_j) = max(-6, log(\frac{N_j}{N} n))$$

The prediction $P_p$ is mapped back to the $[0,1]$ range and
normalized across all actions using the softmax function
$\mbox{softmax}(x)_i = \frac{e^{\frac{x_i}{T}}}{\sum_j
  e^{\frac{x_i}{T}}}$, where $T$ is the temperature parameter that was
set to $2$. The final prior probabilities used in MCTS are

$$P_f(P_t) = \mbox{softmax}(P_t)$$

\section{Experiment Hyperparameters}
\label{app:hyperparameters}
\plc is parameterized with configuration files (see examples in the \texttt{ini} directory of the distributed code), so the key parameters can be easily modified. Here we list the most important hyperparameters used in our experiments.

\paragraph{Feature extraction}
Our main features are term walks of length up to 3. We also add
several more specific features: number of open goals, number of
symbols in them, their maximum size and depth, length of the current
path, and two most frequent symbols in open goals.  Our features are
collapsed (as described in Section~\ref{sec:plcop}) to a fixed 10000-dimensional vector.

\paragraph{MCTS}
MCTS has an inference limit of 200000 steps and an additional time
limit of 200 seconds. Bigsteps are made after 2000 steps. The
exploration constant (cp) is 3 in the first iterations and 2 in later
iterations.

\paragraph{\lc parameters}
\lc usually employs an iteratively increasing path limit to ensure
completeness. We set path limit to 1000, i.e., we practically remove it,
in order to allow exploration at greater depth.

\paragraph{XGBoost parameters}
To train XGBoost models, we use a learning rate of 0.3, maximum tree
depth of 9, a weight decay of 1.5, a limit of 400 training rounds with
early stopping if no improvement takes place over 50 iterations. We
use the built-in ``scale\_pos\_weight'' XGBoost training argument to
ensure that our training data is sign-balanced.

Furthermore, there is an option to keep or filter duplicate inputs
with different target values. Our experiments did not show the importance
of this feature, and all results presented in this paper apply
duplicate filtering.

\end{document}